\PassOptionsToPackage{usenames,dvipsnames,svgnames,table}{xcolor}
\documentclass[conference]{IEEEtran} %
\usepackage{textcomp} %
\usepackage{xcolor}
\usepackage[final]{changes}

\definechangesauthor[name={Remi Dulong}, color=red]{RD}
\colorlet{Changes@Color}{red}

\usepackage{acronym}
\usepackage{amsmath,amssymb}
\usepackage{ifthen}
\usepackage{tikz}
\usepackage{pgfplots}
\usepackage[binary-units]{siunitx}
\usepackage{pifont}
\usepackage{xspace}
\usepackage{multirow}
\usepackage[abbreviations]{foreign}
\usepackage{enumitem}
\usepackage{relsize}
\usepackage{array,booktabs}
\usepackage{algorithm}
\usepackage{listings}
\usepackage[eulergreek]{sansmath}
\usepackage[hyphens]{url}
\usepackage{setspace}%
\usepackage{hyperref}
\usepackage{cleveref}
\usepackage[margin=8pt,skip=5pt,belowskip=0pt,font={small,stretch=1.0},labelfont=bf]{caption}%
\graphicspath{ {figures/} }

\hypersetup{
  colorlinks,%
  citecolor=black,%
  filecolor=black,%
  linkcolor=black,%
  urlcolor=black
}

\ExplSyntaxOn
\cs_set_eq:NN \__siunitx_unit_output_number_sep:
\__siunitx_unit_output_number_sep_aux:
\ExplSyntaxOff

\acrodef{SSD}{solid-state drive}
\acrodef{FUSE}{filesystem in userspace}
\acrodef{NVMM}{non-volatile main memory}
\acrodef{NVM}{non-volatile memory}
\acrodef{DBMS}{database management system}
\acrodef{HDD}{hard disk drive}
\acrodef{DRAM}{dynamic random-access memory}
\acrodef{CDF}{cumulative distribution function}
\acrodef{DAX}{direct access}
\acrodef{CLWB}{cache line write back}
\acrodef{PFENCE}{persistence fence}
\acrodef{API}{application programming interface}
\acrodef{LRU}{least recently used}
\acrodef{OS}{operating system}
\acrodef{PTM}{persistent transactional memory}
 
\xspaceaddexceptions{+}
\xspaceaddexceptions{-}

\newcommand{\sys}{\textsc{NVCache}\xspace}
\newcommand{\systitle}{NVC{\smaller[1.5]ACHE}}

\newcommand{\ramfs}{tmpfs\xspace}
\newcommand{\extdax}{Ext4-DAX\xspace}
\newcommand{\extfour}{Ext4\xspace}
\newcommand{\nova}{NOVA\xspace}
\newcommand{\sqlite}{SQLite\xspace}
\newcommand{\rocksdb}{RocksDB\xspace}

\newcommand{\strata}{Strata\xspace}
\newcommand{\splitfs}{SplitFS\xspace}
\newcommand{\dmwritecache}{DM-WriteCache\xspace}

\newboolean{showcomments}
\setboolean{showcomments}{true}
\ifthenelse{\boolean{showcomments}}
{ \newcommand{\mynote}[3]{
   \fbox{\bfseries\sffamily\scriptsize#1}
        {\small$\blacktriangleright$\textsf{\emph{\color{#3}{#2}}}$\blacktriangleleft$}}
  \newcommand{\zzz}[1]{{\setlength{\fboxsep}{2pt}\fcolorbox{black}{yellow}{\textsf{\emph{#1}}}}\xspace}}
{ \newcommand{\mynote}[3]{}
  \newcommand{\zzz}[1]{}}

\usepgfplotslibrary{external,units,colorbrewer,groupplots}
\usetikzlibrary{patterns}

\makeatletter
\let\origsection\section
\renewcommand\section{\@ifstar{\starsection}{\nostarsection}}
\newcommand\nostarsection[1]{\sectionprelude\origsection{#1}\sectionpostlude}
\newcommand\starsection[1]{\sectionprelude\origsection*{#1}\sectionpostlude}
\newcommand\sectionprelude{\vspace{-6pt}}
\newcommand\sectionpostlude{\vspace{-2pt}}
\let\origsubsection\subsection
\renewcommand\subsection{\@ifstar{\starsubsection}{\nostarsubsection}}
\newcommand\nostarsubsection[1]{\subsectionprelude\origsubsection{#1}\subsectionpostlude}
\newcommand\starsubsection[1]{\subsectionprelude\origsubsection*{#1}\subsectionpostlude}
\newcommand\subsectionprelude{\vspace{-6pt}}
\newcommand\subsectionpostlude{\vspace{-2pt}}
\g@addto@macro\normalsize{%
  \setlength\abovedisplayskip{1pt}
  \setlength\belowdisplayskip{1pt}
  \setlength\abovedisplayshortskip{1pt}
  \setlength\belowdisplayshortskip{1pt}
  \setlength{\floatsep}{2pt}
  \setlength{\textfloatsep}{2pt}
  \setlength{\intextsep}{2pt}
  \setlength{\dblfloatsep}{2pt}
  \setlength{\dbltextfloatsep}{2pt}
}
\makeatother

\newcommand\copyrighttext{%
	\footnotesize \textcopyright 2021 IEEE.
	Personal use of this material is permitted.
	Permission from IEEE must be obtained for all other uses,
	in any current or future media, including reprinting/republishing this
	material for advertising or promotional purposes, creating new collective
	works, for resale or redistribution to servers or
	lists, or reuse of any copyrighted component of this work in other works.
	Pre-print version. Presented in the {51th IEEE/IFIP International Conference on Dependable Systems and Networks (DSN '21)}. For the final published version, refer to DOI \href{https://doi.org/10.1109/DSN48987.2021.00033}{10.1109/DSN48987.2021.00033}}
\newcommand\copyrightnotice{%
\begin{tikzpicture}[remember picture,overlay]
	\node[anchor=south,yshift=10pt,fill=yellow!20] at (current page.south) {\fbox{\parbox{\dimexpr\textwidth-\fboxsep-\fboxrule\relax}{\copyrighttext}}};
\end{tikzpicture}%
}

\begin{document}

\title{NVCache: A Plug-and-Play NVMM-based I/O Booster for Legacy Systems}

\author{\IEEEauthorblockN{Rémi Dulong\IEEEauthorrefmark{1}, Rafael Pires\IEEEauthorrefmark{3}, Andreia Correia\IEEEauthorrefmark{1}, Valerio Schiavoni\IEEEauthorrefmark{1}, Pedro Ramalhete\IEEEauthorrefmark{4},  Pascal Felber\IEEEauthorrefmark{1}, Gaël Thomas\IEEEauthorrefmark{2}}
 \\
 \IEEEauthorblockA{
  \IEEEauthorrefmark{1}Universit\'e de Neuch\^atel, Switzerland, \texttt{first.last@unine.ch}\\
 \IEEEauthorrefmark{3}\textit{Swiss Federal Institute of Technology in Lausanne, Switzerland}, \texttt{rafael.pires@epfl.ch}\\
 \IEEEauthorrefmark{4}Cisco Systems, \texttt{pramalhete@gmail.com}\\
  \IEEEauthorrefmark{2}\textit{Telecom SudParis/Insitut Polytechnique de Paris}, \texttt{gael.thomas@telecom-sudparis.eu}\\
 }
 }

\maketitle
\copyrightnotice

\pagestyle{plain}

\begin{abstract}
This paper introduces \sys, an approach that uses a \ac{NVMM} as a write cache to improve the write performance of legacy applications.
We compare \sys against file systems tailored for \ac{NVMM} (\extdax and \nova) and with I/O-heavy applications
(\sqlite, \rocksdb).
Our evaluation shows that \sys reaches the performance level of the existing state-of-the-art systems for \ac{NVMM},
but without their limitations:
\sys does not limit the size of the stored data to the size of the \ac{NVMM}, and works transparently with unmodified legacy applications, providing additional persistence guarantees even when their source code is not available.

\end{abstract}

\begin{table*}
  \centering
  \footnotesize
  \setlength{\tabcolsep}{5pt}
  \rowcolors{1}{gray!10}{gray!0}
  \caption{\label{table:related} Properties of several NVMM systems, all fully compatible with the POSIX API.}
  \begin{tabular}{l|c|c|c|c|c||c}
	\rowcolor{gray!25}
    & \extdax \cite{corbet2014supporting,wilcox2014add}
    & \nova \cite{xu2016nova}
    & \strata \cite{kwon:17:strata}
    & \splitfs \cite{kadekodi:19:splitfs}
    & \dmwritecache \cite{dmwritecache-icpe19}
    & \sys \\
    \hline
    Offer a large storage space
                                   & $-$            & $-$         & $+$         & $-$          & $+$         & $+$  \\
    Efficient for synchronous durability
                                   & $+$            & $++$        & $++$        & $+$$+$         & $-$         & $+$  \\
    Durable linearizability \cite{izraelevitz2016failure}
                                   & $+$            & $+$         & $+$         & $+$            & $-$        & $+$ \\
    Reuse legacy file systems      & $+$ (\extfour) & $-$         & $-$         & $+$ (\extfour) & $+$ (Any)  & $+$ (Any) \\
    Stock kernel                   & $+$            & $-$         & $-$         & $-$            & $+$        & $+$ \\
    Legacy kernel API
                                   & $+$            & $+$         & $-$         & $-$            & $+$        & $+$  \\
    \hline
  \end{tabular}
\end{table*}

\section{Introduction}
\label{sec:intro}

\Ac{NVMM} is a type of memory that preserves its content upon power loss, is byte-addressable and achieves orders of magnitude better performance than flash memory.
\Ac{NVMM} essentially provides persistence with the performance of a volatile memory~\cite{izraelevitz:19:optane}.
Examples of \ac{NVMM} include phase change memory (PCM)~\cite{burr2010phase,doller2009phase,lee2009architecting,lee2010phase,athmanathan2016multilevel}, resistive RAM (ReRAM)~\cite{akinaga2010resistive}, crossbar RAM~\cite{jo20143d}, memristor~\cite{yang2013memristive} and, more recently, Intel 3D XPoint~\cite{hady2017platform,liu2019initial,micron3dxpoint,intel3dxpoint}.

Over the last few years, several systems have started leveraging \ac{NVMM} to transparently improve input/output (I/O) performance of legacy POSIX applications.
As summarized in~\Cref{table:related}, these systems follow different approaches and offer various trade-offs, each providing specific advantages and drawbacks.
\S\ref{sec:related} details our analysis but, as a first summary, a system that simultaneously offers the following properties does not exist:
\emph{(i)}~a large storage space while using \ac{NVMM} to boost I/O performance;
\emph{(ii)}~ efficient when they provide useful correctness properties such as synchronous durability (\ie, the data is durable when the write call returns) or durable linearizability (\ie, to simplify, a write is visible only when it is durable)~\cite{izraelevitz2016failure}; and
\emph{(iii)}~ easily maintainable and does not add new kernel code and interfaces, which would increase the attack surface of the kernel.

We propose to rethink the design of I/O stacks in order to bring together all the advantages of the previous systems (large storage space, advanced consistency guarantees, stock kernel), while being as efficient as possible.
To achieve this goal, we borrow some ideas from other approaches and reassemble them differently.
First, like \strata~\cite{kwon:17:strata} and \splitfs~\cite{kadekodi:19:splitfs}, we propose to split the implementation of the I/O stack between the kernel and the user space.
However, whereas \strata and \splitfs make the user and the kernel space collaborate tightly, we follow the opposite direction to avoid adding new code and interfaces in the kernel.
Then, as \dmwritecache~\cite{dmwritecache-icpe19} or the hardware-based \ac{NVMM} write cache used by high-end SSDs, we propose to use \ac{NVMM} as a write cache to boost I/Os.
Yet, unlike \dmwritecache that provides a write cache implemented behind the volatile page cache of the kernel and therefore cannot efficiently provide synchronous durability without profound modifications to its code, we implement the write cache directly in user space.

Moving the \ac{NVMM} write cache in user space does, however, raise some major challenges.
The kernel page cache may contain stale pages if a write is added to the \ac{NVMM} write cache in user space and not yet propagated to the kernel.
When multiple processes access the same file, we solve the coherence issue by leveraging the \texttt{flock} and \texttt{close} functions to ensure that all the writes in user space are actually flushed to the kernel when a process unlocks or closes a file.
Inside a process, the problem of coherence also exists if an application writes a part of a file and then reads it.
In this case, the process will not see its own write since the write is only stored in the log and not in the Linux page cache.
We solve this problem by updating the stale pages in case they are read.
Since this reconciliation operation is costly, we use a read cache that keeps data up-to-date for reads.
As the read cache is redundant with the kernel page cache, we can keep it small because it only improves performance in the rare case when a process writes and quickly reads the same part of a file.

As a result, because it combines all the advantages of state-of-the-art systems, our design becomes remarkably simple to deploy and use.
In a nutshell, \sys is a plug-and-play I/O booster implemented only in user space that essentially consists in an \ac{NVMM} write cache.
\sys also implements a small read cache in order to improve the performance when a piece of data in the kernel page cache is stale.
Finally, using legacy kernel interfaces, \sys asynchronously propagates writes to the mass storage with a dedicated thread.
\Cref{table:related} summarizes the advantages of our system.\added[id=, comment=]{ By adding strong persistence guarantees, \sys prevents any rollback effect in case of crash with legacy software, such as DBMS, and obviates the need for a developer to handle data inconsistencies after crashes. Using \sys reduces code complexity without sacrificing performance, and thanks to persistence, the cache layer becomes transparent and reliable, unlike the Linux default RAM page cache.}

Our design provides three main benefits.
First, \sys can easily be ported to any operating system compliant with the POSIX interface:
\sys is only implemented in user space and just assumes that the system library provides basic I/O functions (\texttt{open}, \texttt{pread}, \texttt{pwrite}, \texttt{close}, \texttt{fsync}).\added[id=]{ This design choice only adds the cost of system calls on a non-critical path, while drastically simplifying the overall design.}
Second, by design, \sys offers a durable write cache that propagates the writes to the volatile write cache of the kernel before propagation to the durable mass storage.
While using a volatile write cache behind a durable write cache could seem unexpected, this design has one important advantage: \sys naturally uses the volatile write cache to decrease I/O pressure on the mass storage without adding a single line of code in the kernel.
In details, instead of individually flushing each write that modifies a single page to the mass storage, the kernel naturally combines the writes by updating the modified page in the volatile page cache before flushing the modified page to disk only once.
\strata also combines writes in volatile memory but, because it cannot leverage the kernel page cache by design, \strata must implement combining in its own kernel module.
Finally, \sys naturally benefits from the numerous optimizations provided by the modern stock kernels it builds upon, \eg, arm movements optimization for hard drives~\cite{mumolo2004reducing,akyurek1995adaptive-tocs95} or minimization of write amplification for \ac{SSD}~\cite{sfs-fast12,wisckey-tos17}.

This paper presents the design, implementation and evaluation of \sys.
Using I/O-oriented applications (\sqlite, \rocksdb), we compare the performance of: \sys with a SATA \ac{SSD} formatted in \extfour, a RAM disk (\ramfs), a \ac{DAX} file system backed by a \ac{NVMM} (\extdax), a file system tailored for \ac{NVMM} (\nova), an \ac{SSD} formatted with \extfour boosted by \dmwritecache and a classical \ac{SSD} formatted with \extfour. 
Our evaluation notably shows that:
\begin{itemize}[leftmargin=*,nosep]
  \item Under synchronous writes, \sys reduces by up to 10$\times$ the disk access latency of the applications as compared to an \ac{SSD}, even when using \dmwritecache.
  \item \sys is as fast as (or faster than) \extdax and often as efficient as \nova, but without limiting the working set of the application to the available \ac{NVMM}.
  \item On half of our workloads, \sys remains less efficient than \nova (up to 1.8$\times$).
    We show that this performance limitation results from the reliance on a generic I/O stack to avoid modifications to the kernel.
  \item \sys is implemented in \num{2585}\,LoC only in user space, while offering the same advantages as \strata (large storage space and advanced correctness properties) with its \num{21255}\,LoC, \num{11333} of which are in the kernel.

\end{itemize}

This paper is organized as follows.
We first describe the design of \sys in \S\ref{sec:approach} and detail its implementation in \S\ref{sec:implementation}.
In \S\ref{sec:eval}, we present our evaluation using legacy applications and real-world workloads.
We survey related work in \S\ref{sec:related}, before concluding with some perspectives in \S\ref{sec:conclusion}.

\section{\systitle}
\label{sec:approach}

As outlined in introduction, we have designed \sys with three different goals.
First, we want to use \ac{NVMM} to offer fast \emph{synchronous durability}.
With \sys, instead of routing data writes first to the volatile kernel cache, which is risky as it can lead to data loss upon crash, applications directly and synchronously write to durable storage.
Furthermore, write performance is only limited by the achievable throughput of the actual \ac{NVMM}.

Second, \sys supports legacy applications.
Our design allows applications to persist data at native \ac{NVMM} speed without the size limitations of specialized file systems for \ac{NVMM}, and without requiring any modifications to the application code.

Finally, \sys does not require new kernel interfaces and code, reducing the risk of introducing new attack vectors and simplifying maintainability.
Our design also leverages the numerous existing optimizations implemented in the kernel of the operating system for the storage sub-systems.
In particular, we take advantage of the kernel page cache to increase read performance and to combine writes in memory before propagating them to the mass storage.

\subsection{Approach and workflow}
\label{sec:approach:overview}

\sys implements a write-back cache in \ac{NVMM} and executes entirely in user space.
This design choice avoids kernel modifications and shortens the path between the application and the I/O stack.
Specifically, \sys intercepts the writes of the applications and caches them in a log, stored in \ac{NVMM}.
Then, it asynchronously propagates them to the disk through the kernel using regular I/O system calls. 
Because \sys asynchronously propagates written data to the kernel, the kernel page cache may contain stale data.
For a modified data, \sys has thus to reconstruct a fresh state by applying the last writes recorded in the \sys log during a read operation.
Since this reconciliation operation is costly, \sys also implements a small read cache in user space in volatile memory to keep data up-to-date for reads.

\begin{figure}
	\centering
	\includegraphics[scale=0.8]{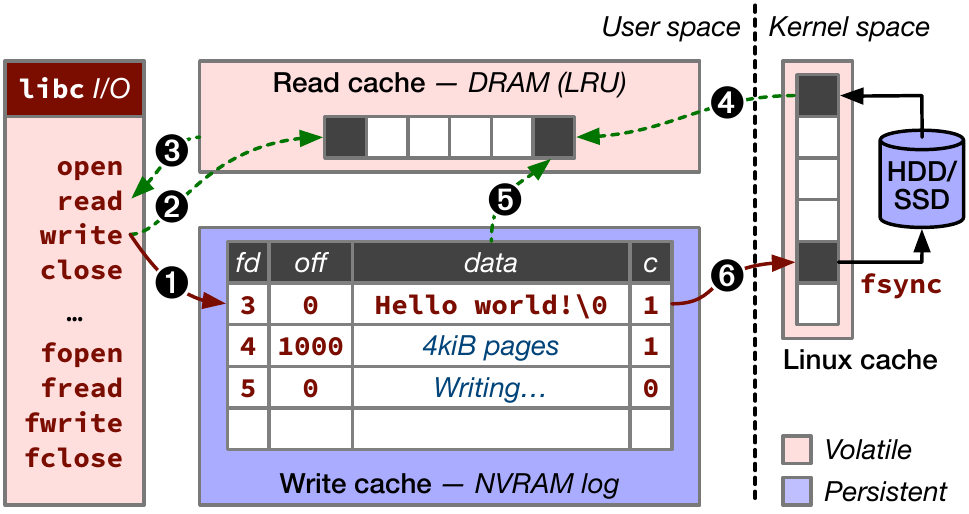}
	\caption{Overall architecture of \label{fig:nvcachemodel}\sys ({\sffamily\itshape fd:} file descriptor, {\sffamily\itshape off:} offset, {\sffamily\itshape c:} committed).}
\end{figure}

\Cref{fig:nvcachemodel} gives a high-level overview of the architecture and workflow.
\sys intercepts application calls to I/O functions of the legacy system library (\texttt{libc}).
In case of a \texttt{write}, \sys adds the written data in the write cache (\Cref{fig:nvcachemodel}-\ding{202}) by appending it to the \ac{NVMM} log, which makes the write durable.

If the write modifies data in the volatile read cache, \sys also updates the data accordingly (\Cref{fig:nvcachemodel}-\ding{203}) so that subsequent reads always see up-to-date data.

In case of a \texttt{read}, \sys retrieves the data from the read cache.
The data is either present and up-to-date (cache hit, \Cref{fig:nvcachemodel}-\ding{204}) or unavailable (cache miss).
Upon cache misses, \sys loads the data from the kernel page cache (\Cref{fig:nvcachemodel}-\ding{205}).
If the data in the kernel is stale, \ie, the data was already modified in the write cache, \sys also applies the writes saved in the \ac{NVMM} log (\Cref{fig:nvcachemodel}-\ding{206}).
 
In  background, a dedicated \emph{cleanup thread} asynchronously propagates writes to the kernel page cache (\Cref{fig:nvcachemodel}-\ding{207}), which will itself subsequently propagate them to disk.
To do so, the cleanup thread uses standard \texttt{write} system calls from the legacy system library.
In addition, the cleanup thread is also in charge of removing pending writes from the log as soon as the kernel has flushed them to disk.

\sys is optimized for applications opening files in read-only mode.
In such instances, \sys entirely bypasses its read caches, hence avoiding the use of \ac{DRAM} altogether, because the kernel page cache already contains fresh data for read-only files.

We first describe the mechanisms underlying the write and read caches assuming a single-threaded application, focusing on multi-threading in \S\ref{sec:multithreading}.

\subsection{Design of the write cache}
\label{sec:nvram:log}

\sys implements its write cache as a circular log in \ac{NVMM}.
Each entry in the log contains a write operation, \ie, the target file descriptor, the starting offset, data itself and the number of bytes to write.

In addition to the log, \sys stores in \ac{NVMM} a table that associates the file path to each file descriptor, in order to retrieve the state after a crash.
\sys also keeps a \emph{persistent tail index} in \ac{NVMM} to track the oldest entry of the log.
The cleanup thread and the recovery procedure use the tail index to propagate the oldest writes to disk before the newer ones,
in order to preserve the write order of the application.

\sys uses a \emph{head index} in volatile memory\footnote{This index is not in \ac{NVMM}, \ie not needed for recovery after a crash.} to create new entries in the log.
When \sys intercepts a write, it advances the head index and fills the new log entry accordingly.
Finally, \sys also maintains a copy of the persistent tail index in volatile memory.
\sys uses this volatile tail index to synchronize a writer with the cleanup thread when the log is full (see \S\ref{sec:cleanup-thread} for details).

\smallskip\noindent\textbf{Failure management.}
When a new entry is created in the log but its fields are not yet completely populated, \ie, it corresponds to a non-committed write, the recovery procedure of \sys ignores the entry.
To detect such scenarios, we could use a \emph{committed index} in \ac{NVMM} to keep track of the last committed entry.
While we considered this approach, we decided to avoid it for three reasons.
First, maintaining a shared index in \ac{NVMM} is inefficient in multi-threaded environments because of the synchronization costs for accessing it consistently.
Second, non-trivial synchronization mechanisms are also required to handle scenarios when a thread commits a new entry before an older one that is not yet committed.
Finally, since data moves from the processor caches to the \ac{NVMM} at cache-line granularity, updating an index would lead to an additional cache miss to load the index.

Instead, in order to handle non-committed writes left in the log after a crash, we directly embed a commit flag with each entry (column {\sffamily\itshape c} in \Cref{fig:nvcachemodel}).
With this design, a thread can independently commit an entry while bypassing the synchronization costs of a shared index.
\sys also avoids an additional cache miss because the commit flag lives inside an entry that already has to be loaded during a write.

\subsection{Design of the read cache}
\label{sec:read-cache}

Since the kernel may contain stale data in its page cache, \sys implements a small read cache in user space.
Technically, \sys implements an approximation of an LRU cache in volatile memory.
When a file is not opened in write-only mode (in which case the read cache is not required), \sys maintains a radix tree~\cite{radixtrees} used to quickly retrieve the pages of the file, an approach similar to \nova~\cite{xu2016nova}.%
\footnote{A page size has to be a power of two because we use a radix tree. Nevertheless, pages in \sys are not related to hardware pages.}

\sys also keeps track (in volatile memory) of the cursor and size of each file as both could be stale inside the kernel, \eg, when a newly appended data block is still in flight in the \sys write cache.

\begin{table}[tb]
\centering
\footnotesize
\renewcommand{\arraystretch}{1.1}
\setlength{\tabcolsep}{3pt}
\caption{\label{tab:states}Page states.}
\begin{tabular}{m{36mm}r|>{\centering\arraybackslash}m{18mm}|>{\centering\arraybackslash}m{18mm}}
\rowcolor{white}
\cellcolor{white}	& \cellcolor{white} &	\multicolumn{2}{c}{\em\textbf{Is page in DRAM read cache?}} \\ 
	&	& \cellcolor{gray!25}\textbf{Yes} & \textbf{No} \\
	\hline
	\cellcolor{white}\em\textbf{Are there corresponding log} & \cellcolor{gray!25}\textbf{Yes} & Loaded & Unloaded dirty \\%
	\cline{2-4}
	\cellcolor{white}\em\textbf{entries in NVMM write cache?} & \textbf{No} & Loaded & Unloaded clean \\%
	\hline
\end{tabular}
\end{table}

\smallskip\noindent\textbf{Page descriptor and page content.}
A leaf of the radix tree contains a \emph{page descriptor}, which is created on demand during a read or a write.
The page descriptor keeps track of the state of a page (see \Cref{tab:states}).
When a page is present in the read cache, the page is in the \emph{loaded} state.
In this state, a \emph{page content} is associated to the page descriptor.
A page content is a piece of cached data, always kept consistent:
when \sys intercepts a write, it also updates the content of a loaded page in the read cache.

When a page is not present in the read cache, its page descriptor (when it exists) is not associated to a page content.
When unloaded, a page can have two different states.
A page is in the \emph{unloaded-dirty} state if the \sys write cache contains entries that modify the page.
In this state, the content of the page outside \sys (in the kernel page cache or on disk) is dirty.
A page is in the \emph{unloaded-clean} state otherwise.
\sys distinguishes the unloaded-dirty from the unloaded-clean state with a counter, called the \emph{dirty counter}, stored in the page descriptor.
\sys increments this counter in case of writes, and the cleanup thread decrements this counter when it propagates an entry from the write cache.

\begin{figure}[!t]
	\centering
	\includegraphics[scale=0.8]{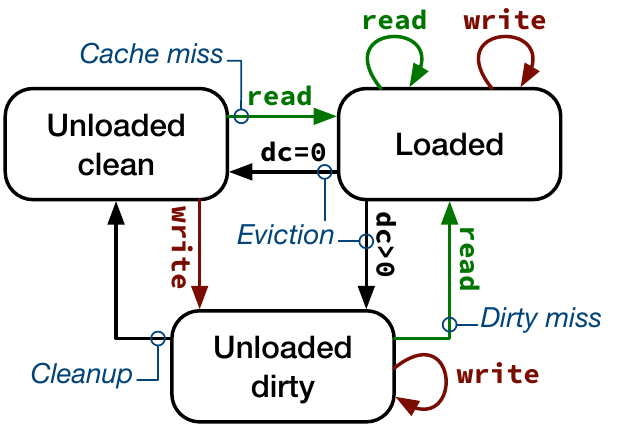}
	\caption{\label{fig:states}State machine of pages ({\sffamily\itshape dc:} dirty counter).}
\end{figure}

\smallskip\noindent\textbf{State transitions.}
Our design has the goal of avoiding any costly synchronous write system call in the critical path of the application.
Technically, \sys avoids write system calls when it evicts a dirty page or when the application modifies an unloaded page.
Instead of synchronously writing the page, \sys simply marks the page as unloaded-dirty and lets the cleanup thread asynchronously perform the propagation.

\Cref{fig:states} presents the state transitions for the pages handled by \sys.
In more detail, in case of cache miss, \sys starts by evicting the least recently used page (loaded to unloaded-clean or to unloaded-dirty according to the dirty counter in \Cref{fig:states}) to make space for the new page.
Then, \sys handles two cases.
If the new page comes from the unloaded-clean state (unloaded-clean to loaded in \Cref{fig:states}), \sys simply loads the page in the read cache by using the \texttt{read} system call.
If the new page comes from the unloaded-dirty state (unloaded-dirty to loaded in \Cref{fig:states}), \sys loads the page and additionally executes a custom \texttt{dirty-miss} procedure.
This procedure reconstructs the state of the page by searching the dirty counter entries that modify the page in the log, starting from the tail index, and applies the modification in the read cache.
This procedure is costly but, as shown in our evaluation (see \S\ref{sec:eval}), dirty misses are rare.
Thanks to the \texttt{dirty-miss} procedure, \sys avoids the synchronous write system calls upon dirty page eviction (loaded to unloaded-dirty in \Cref{fig:states}) and when writing an unloaded page (unloaded-clean to unloaded-dirty in \Cref{fig:states}).

\subsection{Multi-threading}
\label{sec:multithreading}

As required by the POSIX specification, \sys ensures that read and write functions are atomic with respect to each other~\cite{posix-thread-interaction}.%
\sys ensures thus that concurrent writes to the same pages are executed one after the other while respecting the write order of the application, and ensures that a read cannot see a partial write.
Apart from ensuring atomicity for reads and writes on the same pages, the design of \sys also natively supports fully parallel execution of writes to independent pages. 
We achieve this by leveraging three techniques: fixed-sized entries, page-level locking granularity and scalable data structures, which we describe next.

\smallskip\noindent\textbf{Fixed-sized entries.}
As a first step to execute independent writes in parallel, \sys uses fixed-size entries.
With entries of arbitrary size, one would need to store the size of each entry in the log, which prevents a thread to commit an entry if the previous entry is not yet committed.
Indeed, in case of crash, the recovery procedure cannot know if the size of an uncommitted entry is correct, and can thus neither inspect the next entry nor consider it as committed.
With fixed-sized entries, a thread can commit an entry, even if a previous entry allocated by another thread is not yet committed.
In this case, because the entry size is predefined (a system parameter of \sys), the recovery procedure can ignore an uncommitted entry and continue with the next one.

Using fixed-sized entries, \sys must use multiple entries for large writes not fitting in a single entry.
For such large writes, \sys must commit all the entries atomically in order to avoid partial writes.
\sys atomically commits the multiple entries by considering the commit flag of the first entry as the commit flag of the group of entries.
Technically, \sys maintains a \emph{group} index in each entry.
The index is set to -1 for the first entry and, for the following entries, it indicates the position of the first entry.
\sys also saves space, and thus cache misses, by packing the commit flag and the group index in the same integer.

\smallskip\noindent\textbf{Per-page locking.}\label{sec:per-page-lock}
When two threads concurrently modify the same pages, \sys ensures that one write is fully executed before the other.
Instead of using a single write lock for each file, \sys optimizes parallelism by using a per-page locking scheme, in which each page descriptor contains a lock called the \emph{atomic lock}.
In case of a write, a thread starts by acquiring all the atomic locks of the written pages.
Then, the thread adds the write to the log by creating, filling and committing one or multiple entries.
Finally, before releasing the atomic locks, the thread increments the dirty miss counters and, for modified pages in the loaded state, the thread also updates their content in the read cache.

The atomic lock is also used in the read function in order to ensure atomicity.
In case of a read, as for a write, \sys starts by acquiring all the per-page locks, which ensures that a read cannot see a partially updated page.

\sys also uses a second lock per page, called the \emph{cleanup lock}, which is used to synchronize the cleanup thread and the dirty miss procedure.
Without this lock, in case of cache miss, the application may read a stale page from the disk and miss an entry concurrently propagated by the cleanup thread.
We avoid this race condition by acquiring the cleanup lock in the cleanup thread before the propagation, and by acquiring the cleanup lock in the application in case of cache miss.
In more detail, a reader starts by acquiring the atomic lock in order to ensure atomicity.
Then, in case of cache miss, the reader both reads the page from the disk and applies the dirty miss procedure while it owns the cleanup lock, which ensures that the cleanup thread cannot process an entry associated to the page while the application applies the dirty miss procedure.

Finally, \sys synchronizes the access to the dirty miss counter by using atomic instructions and by leveraging the two locks associated to the page.
In case of a write, \sys simply increments the dirty miss counter with an atomic instruction while it owns the atomic lock, which ensures that a reader cannot execute the dirty miss procedure at the same time (because the reader has to own the atomic lock to execute a read and thus the dirty miss procedure).
The cleanup thread decrements the dirty miss counter with an atomic instruction while it owns the cleanup lock, which ensures that a reader cannot execute the dirty miss procedure at the same time (because the reader has also to own the cleanup lock to execute the dirty miss procedure).

Due to our locking scheme with two locks per page, the cleanup thread never blocks a writer and never blocks a reader when the page is already in the read cache.
The cleanup thread can only block a reader in case of cache miss when the cleanup thread propagates an entry that modifies the missed page.

\smallskip\noindent\textbf{Scalable data structures.}
Most of the internal structures of \sys are designed to scale with the number of threads.
First, in order to add an entry in the write log, a thread simply has to increment the head index, which is done with atomic instructions.%
\footnote{Note that we do not need special instructions to handle \ac{NVMM} since the head lives in volatile memory.}

Then, the radix tree can operate with a simple lock-free algorithm because \sys never removes elements from the tree, except when it frees the tree upon close.
When an internal node or a page descriptor in the radix tree is missing, a thread tries to atomically create it with a compare-and-swap instruction; if this fails, it means that the missing node or page descriptor was concurrently added to the tree by another thread and we can simply use it.

Finally, \sys builds an approximation of an LRU algorithm by storing an \emph{accessed} flag in each page descriptor, which is set during a read or a write.
In more detail, \sys maintains a queue of page contents protected by a lock, called the \emph{LRU lock}.
Each page content holds a reference to its descriptor and conversely, as already presented, a descriptor links to the associated page content when it is loaded.
When \sys has to evict a page, it acquires the LRU lock and dequeues the page content at the head of the queue.
Then, \sys acquires the atomic lock of the page descriptor associated to the head and checks its accessed flag.
If the flag is set, \sys considers that the page was recently accessed:
it releases the atomic lock of the page descriptor, re-enqueues the page content at the tail, and restarts with the new head.
Otherwise, \sys recycles the page content: it nullifies the reference to the page content in the page descriptor, which makes the page descriptor unloaded-clean or unloaded-dirty, and releases the atomic lock of the page descriptor.

\smallskip\noindent\textbf{Summary.}
To summarize, in case of a write, \sys only acquires the atomic locks to the written pages.
In case of read hits, \ie, if the page contents are already in the read cache, \sys also only acquires the atomic locks to the read pages.
In case of read misses, \sys has also to acquire the LRU lock and an atomic-lock during eviction, and then the cleanup lock to read the missed page.
As a result, except to handle read misses, \sys executes two operations that access different pages in a fully concurrent manner, and without using locks to synchronize with the cleanup thread.

\section{Implementation}
\label{sec:implementation}
We implemented \sys on top of the \texttt{musl} libc~\cite{musl:20}, a lightweight yet complete alternative to \texttt{glibc}.
\texttt{Musl} consists of \num{85}\,kLoC and \sys adds \num{2.6}\,kLoC.
In order to simplify deployment, instead of using \texttt{LD\_PRELOAD} to wrap I/O-based system calls, we directly replace the I/O-based system calls by ours in \texttt{musl}.
We also exploit the Linux Alpine distribution~\cite{alpine:20} using \texttt{musl} to easily deploy \sys behind legacy applications (as shown in our evaluation).
Moreover, \sys supports Docker containers~\cite{docker:20} and our approach allows us to deploy legacy applications with just one minor change to existing manifests, replacing the original libc shared object by ours.

\begin{table}[!t]
\centering
\footnotesize
\setlength{\tabcolsep}{3pt}
\caption{\label{tab:functions}Functions intercepted by \sys.}
\rowcolors{1}{gray!10}{gray!0}
  \begin{tabular}{r|l}
  \rowcolor{gray!25}
    \textbf{Function} & \textbf{Action} \\
    \hline
    \texttt{open}, \texttt{read}, \texttt{write}, \texttt{close}     & Uses \sys functions \\
    \texttt{fopen}, \texttt{fread}, \texttt{fwrite}, \texttt{fclose} & Uses unbuffered versions \\
    \texttt{sync}, \texttt{syncfs}, \texttt{fsync}                   & No operation \\
    \texttt{lseek}, \texttt{ftell}, \texttt{stat}, \etc.             & Uses size/cursor of \sys \\
    \hline
  \end{tabular}
\end{table}

\Cref{tab:functions} lists the main functions intercepted by \sys.
Essentially, \sys wraps \texttt{open}, \texttt{read}, \texttt{write} and \texttt{close} to use the read and write caches.
Additionally, it replaces the buffered versions of I/O functions (\texttt{fread}, \texttt{fwrite}, \texttt{fopen}, \texttt{fclose}) with their unbuffered counterparts, since \sys itself acts as a buffer in user-space with its small read cache in volatile memory.

As for the \texttt{fsync} function calls, which force the kernel to synchronously propagate pending writes from the kernel page cache to the disk, \sys simply ignores them.
These operations are no longer necessary because \sys already makes the write synchronously durable.

As presented in \S\ref{sec:read-cache}, \sys maintains its own versions of file cursors and size information, because the cursor associated to an opened file and the size of a file may be stale in the kernel. %
Therefore, \sys intercepts the \texttt{stat} and \texttt{seek} function families in order to return fresh values.

At low level, in order to read and write from a cursor maintained by \sys and not the one maintained by the kernel, the cleanup thread uses the \texttt{pwrite} function to propagate an entry to the kernel and, upon cache misses, \sys uses \texttt{pread} to load a page in the read cache.

\added[id=, comment=]{\sys does not support asynchronous writes, but they could be implemented. Memory-mapped files, however, are not supported in \sys and their implementation remains an open problem, as loads and stores are not interceptable at the libc level.}

In the following, we detail how we have implemented the \texttt{open}, \texttt{write}, \texttt{cleanup} and \texttt{recovery} functions because they directly deal with \ac{NVMM}.
The \texttt{read} function closely implements the design presented in \S\ref{sec:read-cache} and \S\ref{sec:multithreading}. 

\smallskip\noindent\textbf{Open.}
The \texttt{open} function adds a file to the cache.
First, in \ac{NVMM}, \sys maintains a table that associates the paths of the opened files to their file descriptors.
This table is only used during recovery:
after a crash, the file descriptors in the log entries are meaningless and the recovery engine therefore needs the information in the table to retrieve the paths of the files associated to these file descriptors.

Then, in volatile memory, \sys keeps track of opened files with two tables in order to handle independent cursors when an application opens the same file twice.
The first one, called the file table, associates a \emph{(device, inode number)} pair to a file structure, which contains the size of the file and its radix tree.
The second one, called the opened table, associates a file descriptor to an opened-file structure, which contains the cursor and a pointer to the file structure.

In \texttt{open}, \sys starts by retrieving the device and the inode number associated to the file path with the \texttt{stat} system call.
Then, if the file belongs to a block device and if the \emph{(device, inode number)} pair is not yet present in the file table,
\sys creates the file structure.
Finally, \sys uses the \texttt{open} system call to create a file descriptor and creates accordingly an opened-file structure in the opened table.

\sys bypasses its read cache when a file is only opened in read-only mode.
For that purpose, \sys only creates a radix tree in the file structure when the file is opened in write mode for the first time and, for a file that does not have a radix tree, \sys bypasses its read cache.

\begin{algorithm}[t!]
\caption{--- \sys \texttt{write} function.}\label{code:write}
\lstset{
basicstyle=\linespread{1.0}\footnotesize\sffamily,
commentstyle=\color{RoyalBlue}\footnotesize\sffamily\slshape,
columns=fullflexible,
language=C++,
numbers=left,
stepnumber=1,
numberstyle=\tiny\sffamily,
numbersep=3pt,
xleftmargin=8pt,
numberblanklines=false,
escapeinside={(*@}{@*)}
}
\begin{lstlisting}
struct nvram {          (*@\hfill@*) (*@{\color{BrickRed}\slshape\bfseries // Non-volatile memory}@*)
  struct { char path[PATH_MAX]; } fds[FD_MAX];
  struct entry entries[NB_ENTRIES];
  uint64_t persistent_tail; 
}* nvram;

uint64_t head, volatile_tail; (*@\hfill@*) (*@{\color{BrickRed}\slshape\bfseries // Volatile memory}@*)

void write(int fd, const char* buf, size_t n) {
  struct open_file* o = open_files[fd];
  struct file* f = o->file;
  struct page_desc* p = get(f->radix, o->offset);

  uint64_t index = next_entry();
  struct entry* e = &nvram->entries[index % NB_ENTRIES];
  
  acquire(&p->atomic_lock);

  memcpy(e->data, buf, n);    (*@\hfill@*) (*@{\color{BrickRed}\slshape\bfseries // Write cache}@*)
  e->fd = fd;
  e->off = o->off;
  pwb_range(e, sizeof(*e)); (*@\hfill@*) // Send the uncommited entry to NVMM
  pfence();              (*@\hfill@*) // Ensure commit is executed after

  e->commit = 1;                     
  pwb_range(e, CACHE_LINE_SIZE);          (*@\hfill@*) // Send the commit to NVMM
  psync();                   (*@\hfill@*) // Ensure durable linearizability

  atomic_fetch_add(&p->dirty_counter, 1); (*@\hfill@*) (*@{\color{BrickRed}\slshape\bfseries // Read cache}@*)
  if(p->content)                (*@\hfill@*) // Update page if present in the read cache
    memcpy(p->content->data + o->off %
  release(&p->atomic_lock);
}

int next_entry() {
  int index = atomic_load(&head);
  while(((index + 1) %
        !atomic_compare_and_swap(&head, index, index + 1))
    index = atomic_load(&head);
    return index; (*@\hfill@*) // Commit flag at index is 0 (see cleanup thread)
}
\end{lstlisting}
\end{algorithm}

\smallskip\noindent\textbf{Write.}
\Cref{code:write} shows a simplified and unoptimized version of \sys's write function when the write fits in one page and one entry.
Our code uses three \ac{NVMM}-specific instructions: \texttt{pwb(addr)} (\eg, \texttt{clwb} on a Pentium)
ensures that the cache line that contains \texttt{addr} is added in the flush queue of the cache;
\texttt{pfence} (e.g., \texttt{sfence} on a Pentium) acts both as a store barrier and ensures that the \texttt{pwbs} that precede are executed before the barrier;
and \texttt{psync} (e.g. also \texttt{sfence} on a Pentium) acts as a \texttt{pfence} and furthermore ensures that the cache line is actually drained to the \ac{NVMM}.
With these instructions, the \texttt{write(a, v1)}, \texttt{pwb(a)}, \texttt{pfence}, \texttt{write(b, v2)} sequence ensures that the write to \texttt{a} is propagated to \ac{NVMM} before the write to \texttt{b} because the cache line of \texttt{a} is flushed before the write to \texttt{b} is enqueued in the flush queue of the cache.

In our code, after having retrieved (or lazily created) the page descriptor (line~12), the write function finds a new free entry in the log (line~14 and~35--41).
In details, \texttt{next\_entry} first waits if the log is full (line~37) and then advances the head while taking care of concurrent accesses from another threads (line 38).
At this step, the commit flag of the entry returned by \texttt{next\_entry} is necessarily equal to 0 (see the cleanup thread below).

Then, as soon as the write function acquires the lock of the page descriptor (line~17), it adds the write to the log (lines~19 to~27).
More precisely, the function fills the entry without committing it (lines~19 to~21), sends the uncommitted entry to the \ac{NVMM} by flushing the cache lines of the entry (line~22) and executes a \texttt{pfence} in order to ensure that the entry is flushed before the commit (line~23).
The function then commits the entry (line~25), sends the cache line that holds the commit flag to the \ac{NVMM} (line~26) and executes a \texttt{psync} to ensure durable linearizability (line~27, see the text below for the explanation).
\added[id=]{At this level, the \texttt{atomic\_lock} is only taken for atomicity purposes between the writer thread and the cleanup thread, preventing the cleanup thread to modify a page on the SSD while a cache miss procedure reads it.}

Finally, the write function manages the read cache (lines~29 to~31) before releasing the lock at line~32.
Specifically, it increments the dirty counter (line~29) because the log contains a new entry that modifies the page.
The increment is done with an atomic instruction in order to prevent a conflict with the cleanup thread that may be concurrently decrementing the counter (see \S\ref{sec:multithreading}).%
\footnote{Because a writer does not acquire the cleanup lock, the cleanup thread may decrement the dirty counter between the commit at line~27 and the atomic add at line~29, making the counter negative.
This temporary negative counter cannot lead to an incorrect behavior because a reader has to take both the atomic and cleanup locks to execute the dirty miss procedure, ensuring that the dirty miss procedure cannot observe an unstable negative counter.}
If the page is in the loaded state (\ie, the page descriptor is associated to a page content, line~30), the function updates the page content in the read cache (line~31).

\smallskip\noindent\textbf{Durable linearizability.}
Our algorithms ensure durable linearizability~\cite{izraelevitz2016linearizability}, which essentially means that if a read sees a write, then the write is committed.
For example, this is not the case of Linux when it uses the page cache:
indeed, a thread can see a write stored in the page cache but not yet propagated to disk.
\sys always ensures durable linearizability because of the \texttt{psync} at line~27.
This operation ensures that the commit at line~25 is written to \ac{NVMM} before the lock release at line~32, which is itself executed before the lock acquire of a reader able to see the write.

\smallskip\noindent\textbf{Cleanup thread and batching.}
\label{sec:cleanup-thread}
The cleanup thread propagates the writes from the \ac{NVMM} log to the disk.
At high level, it consumes the entries one by one, starting at the persistent tail index.
The cleanup thread begins by synchronizing with the application through the commit flag:
if the entry at the persistent tail is not yet committed, the cleanup thread waits.
When the entry at the persistent tail is committed, the cleanup thread consumes the entry while owning the cleanup locks associated to the descriptors of the pages modified by the entry.
In more detail, the cleanup thread proceeds in three steps when it consumes an entry.
During the first step, the cleanup thread propagates the entry to the mass storage by using \texttt{pwrite} to send the write to the kernel page cache and by using \texttt{fsync} to synchronously propagate the writes
from the kernel page cache to the mass storage.
During the second step, the cleanup thread updates both the commit flag of the consumed entry and the persistent tail index, and uses \texttt{pwb}/\texttt{pfence} to ensure that the third step can only start after the second step.
During the third step, the cleanup thread marks the entry as free for the writers by using the volatile tail index.
Because of the use of the two tail indexes, when a writer sees that an entry is free in volatile memory (volatile tail index), we have the guarantee that the entry is also marked as free in \ac{NVMM} (persistent tail index and commit flag of the entry).

The implementation described above is inefficient because a call to \texttt{fsync} is especially costly.
The throughput of a random \SI{4}{\kilo\byte} write on an SSD is at least 13$\times$ faster without \texttt{fsync}~\cite{10.1145/2588555.2595632}.
To mitigate the negative impact of a slow cleanup thread that continuously calls \texttt{fsync}, which would otherwise block all the writes of the application when the log is full, the cleanup thread batches the writes.
Batching allows us to reduce the frequency of calls to \texttt{fsync}, updating the tail index only upon success.
The advantages of batching are twofold.
First, batching decreases the number of calls to \texttt{fsync}, boosting the performance of the cleanup thread
and thus decreasing the probability of having a full log.
Then, batching allows \sys to leverage kernel optimizations by combining writes or optimizing the sequences of writes for hard drives or SSD.
\S\ref{sec:batching} presents in detail how batching improves performance.

\smallskip\noindent\textbf{Recovery procedure.}
When \sys starts, it executes a recovery procedure.
It first re-opens the files by using the table that associates file paths to file descriptors stored in \ac{NVMM}.
Then, it propagates all the committed entries of the log by starting at the tail index, invokes the \texttt{sync} system call to ensure that the entries are written to disk, close the files and empties the log.

\added[id=, comment=]{\smallskip\noindent\textbf{Multi-application.}
The NVMM write log of \sys is either a DAX device, \eg, an entire NVMM module, or a DAX file, \ie, a file in any DAX-capable filesystem. Therefore, in a multi-application context, two instances of NVcache can run simultaneously on the same machine, either with one NVMM module each or sharing the same module split into two DAX files.}

\section{Evaluation}
\label{sec:eval}

\subsection{Hardware and software settings}

We evaluate \sys on a Supermicro dual socket machine with two NUMA domains.
Each NUMA domain contains an Intel Xeon Gold 5215 CPU at \SI{2.50}{\giga\hertz} with 10 cores (20 hardware threads), 6$\times$\SI{32}{\gibi\byte} of DDR4 and 2$\times$\SI{128}{\gibi\byte} of Optane NVDIMM.
In total, the machine contains 20 cores (40 hardware threads), \SI{384}{\gibi\byte} of DDR4 and \SI{512}{\gibi\byte} of Optane NVDIMM.
The machine is equipped with two SATA Intel SSD DC S4600 with \SI{480}{\giga\byte} of disk space each.
One of them contains the whole system, while the other is dedicated for some of our experiments.
The main SSD is formatted with an \extfour file system.
The secondary one is mounted with \texttt{lvm2}, linked with a \dmwritecache stored in one of our Optane NVDIMM.
This virtual \texttt{lvm2} device is also formatted with \extfour.
We deploy Ubuntu 20.04 with Linux version 5.1.0 (\nova~\cite{xu2016nova} repository version) and \texttt{musl} v1.1.24, revision \texttt{9b2921be}.

\begin{table*}[!t]
\centering
\footnotesize
\rowcolors{1}{gray!10}{gray!0}
\caption{\label{tab:eval}Evaluated file systems.}
\begin{tabular}{c|c|c|c|c|c}%
  \rowcolor{gray!25}
  \textbf{Name} & \textbf{Write cache} & \textbf{Storage space} & \textbf{FS}  & \textbf{Synchronous durability} & \textbf{Durable linearizability} \\
  \hline
  \sys+SSD            & \sys                        & SSD       & \extfour  & by default & by default \\
  \dmwritecache       & kernel page cache           & SSD       & \extfour  & O\_DIRECT | O\_SYNC    & no \\
  \extdax             & kernel page cache           & \ac{NVMM} & \extfour  & O\_DIRECT | O\_SYNC    & no \\
  \nova\footnotemark
                      & none                        & \ac{NVMM} & \nova     & O\_DIRECT | O\_SYNC & by default \\
  SSD                 & kernel page cache           & SSD       & \extfour  & O\_DIRECT | O\_SYNC    & no \\
  \ramfs              & kernel page cache           & DDR4      & none      & no         & no \\
  \sys+\nova          & \sys                        & \ac{NVMM} & \nova     & by default & by default \\
  \hline
\end{tabular}
\end{table*}

We evaluate \sys with representative benchmarks that heavily rely on persistent storage. 
\rocksdb is a persistent key-value store based on a log-structured merge tree widely used in web applications and production systems, \eg, by Facebook, Yahoo! and LinkedIn~\cite{cao20rockdbfast20}.
We evaluate \rocksdb v6.8 with the \texttt{db\_bench} workload shipped with LevelDB, an ancestor of \rocksdb, which stresses different parts of the database.
\sqlite~\cite{sqlite} is a self-contained SQL database widely used in embedded systems, smartphones and web browsers~\cite{jeong2013stack,kim2012revisiting}.
We evaluate \sqlite v3.25 with a port of the \texttt{db\_bench} for \sqlite.
We also use FIO~\cite{axboe2005fio} version 3.20, a micro-benchmark designed to control the read and write ratios and frequencies, the number of threads or the read and write patterns (random, sequential).

Unless stated otherwise, we configure \sys as follow\added[id=, comment=]{s}.
Each entry in our NVM log is \SI{4}{\kibi\byte} large.
The log itself is constituted of \SI{16}{million} entries (around \SI{64}{\gibi\byte}).
The RAM cache uses \SI{250}{thousand} pages of \SI{4}{\kibi\byte} each (around \SI{1}{\gibi\byte}).
The minimum number of entries before attempting to batch data to the disk is \SI{1}{thousand}.
The maximum number of entries in a batch is \SI{10}{thousand}.

\subsection{Comparison with other systems}
\label{sec:macro}

In this experiment, we compare \sys with other systems.
\Cref{tab:eval} summarizes the different file systems evaluated.
We compare the normal version of \sys when it propagates the writes to an SSD formatted in \extfour (\sys+SSD) with five other systems and a variant of \sys.

\footnotetext[5]{Formally, \nova provides synchronous durability and durable linearizability when mounted with the \texttt{cow\_data} flag.}
\footnotetext[6]{https://ext4.wiki.kernel.org/index.php/Clarifying\_Direct\_IO\%27s\_Semantics}

Specifically, we evaluate:
\emph{(i)}~\dmwritecache, which asynchronously propagates the writes from the volatile kernel page cache to an \ac{NVMM} write cache, and only later propagates the writes from \ac{NVMM} to an SSD formatted in \extfour (\dmwritecache+SSD);
\emph{(ii)}~the \extfour file system directly stored in \ac{NVMM} (\extdax);
\emph{(iii)}~the \nova file system~\cite{xu2016nova}, tailored to efficiently use \ac{NVMM} (\nova),
\emph{(iv)}~an SSD formatted in \extfour (SSD);
\emph{(v)}~a temporary file system, which only stores the data in volatile memory in the kernel page cache (\ramfs).

We also evaluate a variant of \sys that propagates the writes to the \ac{NVMM} formatted with the \nova file system (\sys+\nova).
This variant does not offer a large storage space like \sys+SSD but shows the theoretical performance that we could expect from \sys when using a\added[id=, comment=]{n} efficient secondary storage.

\begin{figure*}
  \centering
  \includegraphics{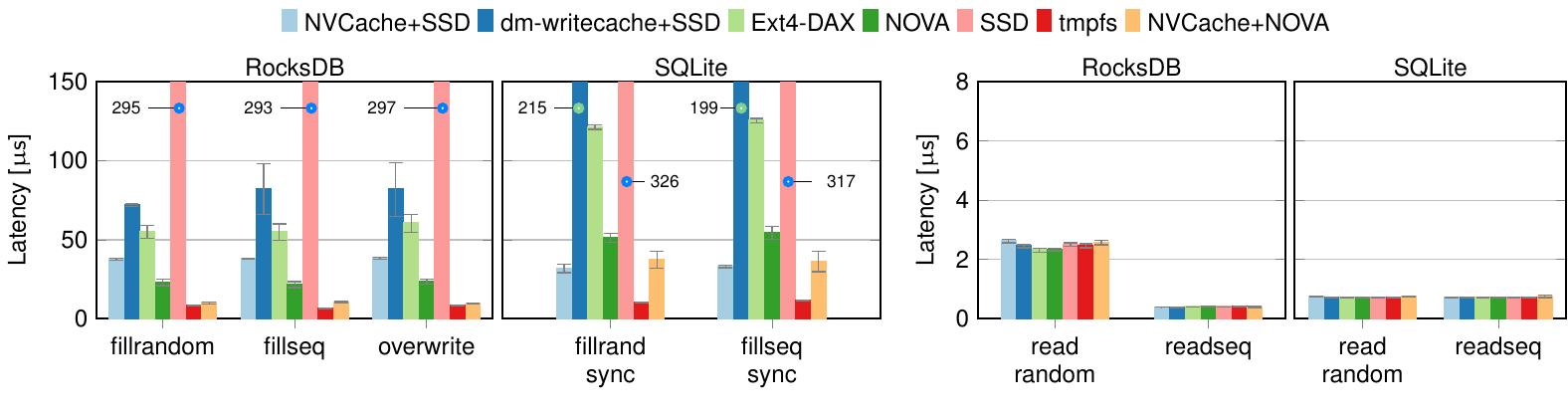}
  \caption{\label{fig:all:write-sync}Performance of \sys for synchronous write-heavy (left) and read-heavy (right) workloads.}
\end{figure*}

\sys+SSD, \nova and \sys+\nova provide the highest consistency guaranties since they offer both synchronous durability (\ie, the data is durable when the write call returns) and durable linearizability (essentially,
a write is only visible when it is durable).
In order to make a fair comparison, we also enforce synchronous durability for all the file systems by \added[id=, comment=]{activating the synchronous mode of our benchmarks. Alternatively, on a non-synchronous benchmark, we could }open
\deleted[id=]{ing} 
the files with the \texttt{O\_SYNC} flag, which guaranties that a write is flushed to disk when the system call returns.
We \added[id=]{can }also optimize these systems by using the \texttt{O\_DIRECT} flag\footnotemark, which tries to avoid an in-memory copy of the buffer from the user space to the kernel space when possible.
\dmwritecache+SSD, SSD and \extdax are not designed to ensure durable linearizability and do not offer the guarantee that a write is visible only when it is durable.
The \ramfs file system does not provide durability, and thus no consistency guaranty.
\Cref{fig:all:write-sync} presents our results, respectively for write-oriented (left) and read-oriented (right) workloads.

\begin{figure*}
  \centering
  \includegraphics{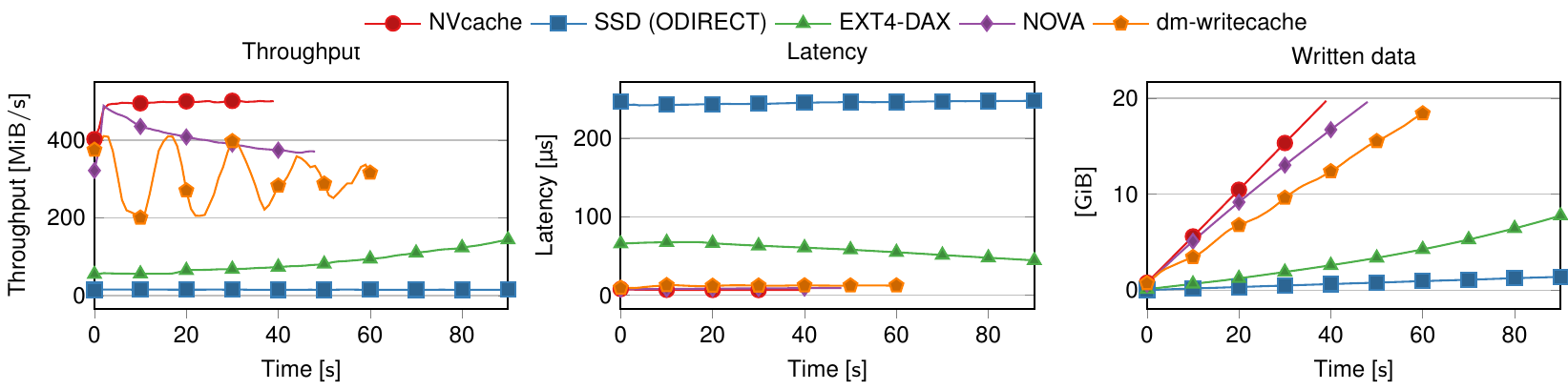}
  \vspace*{-2mm}
  \caption{\label{fig:all:fio-traces} Performance of \sys under random write intensive loads for \SI{20}{\gibi\byte}.}
\end{figure*}

\textbf{Read-oriented workloads.}
Analyzing the read-oriented workloads, we note how all the systems provide roughly the same performance.
This indicates that the different designs do not significantly change the read performance on these benchmarks.
Despite \nova and \extfour are reading from \ac{NVMM} while all the others are reading from an SSD, they all benefit from a volatile read cache stored in DDR4 RAM.

\textbf{Write-oriented workloads.}
When we analyze the write-oriented workloads, we observe that the different designs approaches have a large impact on performance.
We first turn our attention to the systems that offer a large storage space: \sys+SSD, \dmwritecache+SSD and SSD.
We observe that \sys+SSD is consistently faster than the other systems (at least 1.9$\times$).
SSD has the worst performance as it does not leverage \ac{NVMM} to boost performance.
This is not the case of \dmwritecache+SSD and \sys+SSD, both using \ac{NVMM} as a write cache to boost I/O performance.
However, \sys+SSD performs significantly better than \dmwritecache+SSD.
Indeed, the design of \sys+SSD naturally offers synchronous durability, because the application writes directly in the \ac{NVMM}.
Since, by design, \dmwritecache+SSD lives behind the volatile kernel page cache, enforcing synchronous durability requires the execution of additional code during a write.
This code significantly hampers the performance of \dmwritecache+SSD.
Among the three systems, \sys+SSD is also the only system that ensures durable linearizability.

We now focus on systems that offer strong correctness guarantees but sacrifice the storage space: \extfour and \nova.
With \rocksdb, \sys+SSD is 1.4$\times$ faster than \extfour, and \nova is 1.6$\times$ better than \sys+SSD.
With \sqlite, \sys performs better than \nova (around 1.6$\times$ better), and \sys is roughly 3.7$\times$ better than \extfour.
For some workloads, \nova is more efficient than the other systems because it was specifically tailored for \ac{NVMM} and bypasses the bottlenecks of \extfour~\cite{xu2016nova}.
With \rocksdb, we observe that \sys also suffers from these bottlenecks.
Indeed, when we use \sys as an I/O booster in front of \nova instead of SSD (\sys+\nova), \sys can match and even improve performance as compared to \nova.
Overall, these results show that \sys is able to reach performance comparable to a generic file system on \ac{NVMM}, while ignoring the limit of \ac{NVMM} storage space.
Our design remains, however, less efficient than a file system specifically tailored for \ac{NVMM} on some of the workloads because it remains totally independent from the kernel I/O stack.

\subsection{Analysis of \sys}
\label{sec:micro}

In this section, we analyse the behavior of \sys by using the FIO profiling tool.
We configured FIO with the \texttt{fsync=1} flag to ensure that a write is synchronously durable, and with the \texttt{direct=1} flag to open all the files in direct I/O mode.
We set the buffer size to \SI{4}{\kibi\byte} with \texttt{ioengine=psync}.
FIO measures are fetched every second.

\textbf{Comparative behavior.}
\label{eval:comparative}
The objective of this experiment is to study the performance of \sys in an ideal case.
We configure FIO to generate a random write intensive workload and, in \sys, we use a log of \SI{32}{\gibi\byte} for \SI{20}{\gibi\byte} of written data.
As a result, \sys cannot saturate the log and is never slowed down by the cleanup thread.

\Cref{fig:all:fio-traces} reports the performance with these settings.
The left graph shows the instantaneous throughput as it evolves during the run.
The middle graph shows how the average latency evolves during the run.
Finally, the right graph shows how the total amount of written data evolves during the run.
For these measures, we split the run in small periods and report the average throughput observed during each of them (instantaneous throughput), as well as the average latency and cumulative data written as measured from the beginning of the run to the end of each period.

\Cref{fig:all:fio-traces}~(left) shows that, in this ideal case, \sys has a better throughout than all the other systems.
\sys significantly outperforms SSD, \extdax and \dmwritecache (at least 1.5$\times$).
These systems are designed to leverage the volatile Linux page cache in order to improve read performance, which makes them inefficient for writes when an application requires strong consistency guaranties such as synchronous durability.
We can observe that \sys also outperforms \nova on this benchmark (\SI{493}{\mebi\byte\per\second} vs. \SI{403}{\mebi\byte\per\second} on average).
\sys is slightly more efficient than \nova in this ideal case because \sys never calls the system during a write, whereas \nova has to pay the cost of system calls on the critical path.
In \Cref{fig:all:fio-traces}~(middle) and \Cref{fig:all:fio-traces}~(right), we can also observe that the average latency and the written data are better in \sys than in other systems.
As a result, in this ideal case, \sys writes all the data in \SI{42}{\second}, while it takes \SI{51}{\second} for \nova, \SI{71}{\second} for \dmwritecache+SSD, \SI{2}{\minute} and \SI{29}{\second} for \extfour and more than \SI{22}{\minute} for SSD.

\begin{figure*}
    \centering
    \includegraphics{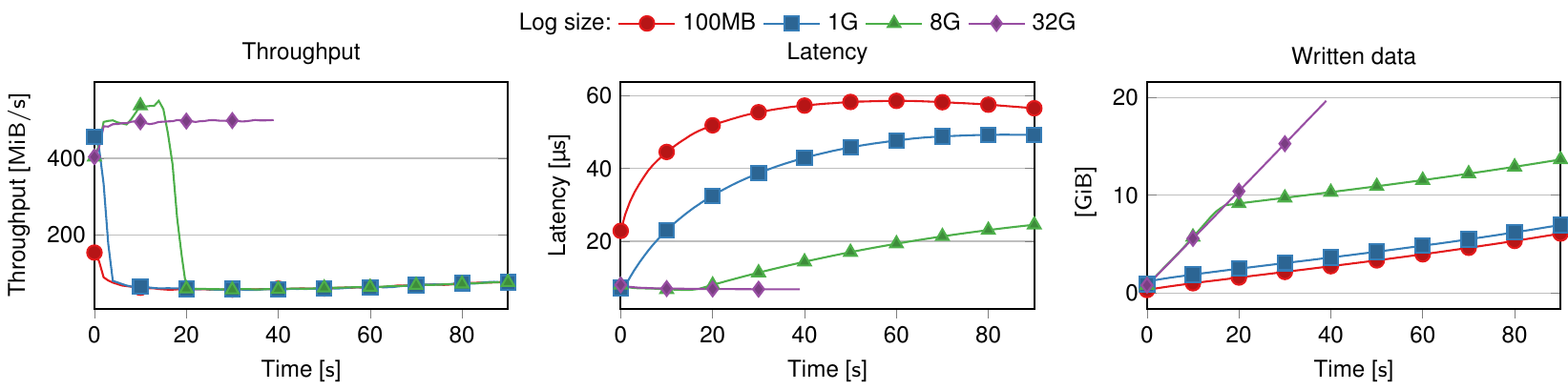}
    \caption{\label{fig:all:fio-logsize} Performance of \sys under random write intensive loads for \SI{20}{\gibi\byte}, with variable \ac{NVMM} log size.}
\end{figure*}

\begin{figure*}
  \centering
  \includegraphics{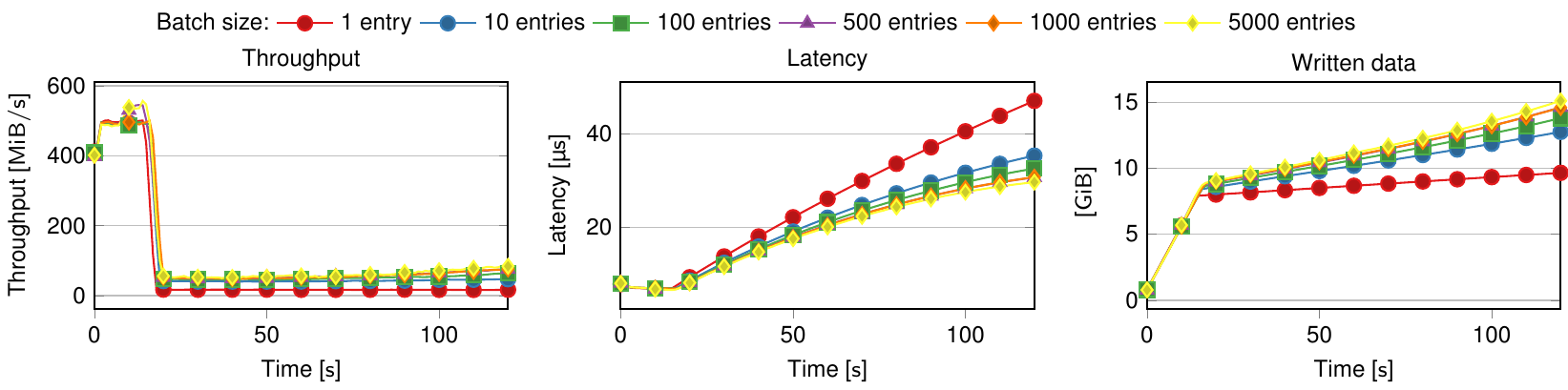}
  \caption{\label{fig:all:fio-batchsize} Influence of batching and batch size parameter.}
\end{figure*}

\textbf{Log saturation.}\label{eval:log-saturatation}
During a long run that intensively writes, \sys may saturate its log, because the cleanup thread can only propagate the writes from the log to the SSD at the speed of the SSD, which is much slower than \ac{NVMM}.
The next experiment highlights this behavior.
Using a workload to intensively write \SI{20}{\gibi\byte} at random locations, we measure the performance of \sys with different log sizes.

\Cref{fig:all:fio-logsize} reports the result of \sys with these settings (instantaneous throughput on the left, average latency in the middle and cumulative data written on the right). 
With a log of \SI{32}{\gibi\byte} and \SI{20}{\gibi\byte} of written data, the log never saturates.
Hence, as in \Cref{fig:all:fio-traces} (see \S\ref{eval:comparative}), the instantaneous throughput and the average latency remain stable.

As shown in \Cref{fig:all:fio-logsize}~(left), we can observe two phases with a log of \SI{8}{\gibi\byte}.
During a first phase that starts at 0 and ends at \SI{18}{\second}, the throughput remains stable and high (\SI{556}{\mebi\byte\per\second} on average). %
At the end of this phase, the throughput suddenly collapses to \SI{78}{\mebi\byte\per\second} and then remains stable until the end of the run.
In this experiment \sys collapses at \SI{18}{\second} because of the log saturation.
Still, before saturation, \sys is as fast as with a log of \SI{32}{\gibi\byte} and only limited by the performance of the \ac{NVMM}.
During the first phase, \sys fills the log but the cleanup thread cannot empty it fast enough because the cleanup thread is limited by the performance of the SSD.
After saturation, FIO has to wait for the cleanup thread, which limits performance to the speed of the SSD.

We can observe a similar pattern for the average latency in \Cref{fig:all:fio-logsize}~(middle).
The latency remains stable and high before saturation and then starts degrading (note that the latency does not collapse because the figure reports the average latency since the beginning of the run).
We can also observe that the slope of throughput curves change when the log saturates in \Cref{fig:all:fio-logsize}~(right): when reaching saturation, the number of written data increases much slower.
We can observe exactly the same behavior with smaller log sizes, but the log saturates earlier.
Interestingly, with a log of \SI{100}{\mebi\byte}, \SI{1}{\gibi\byte} and \SI{8}{\gibi\byte}, the write throughput becomes identical as soon as the log saturates, staying at around \SI{80}{\mebi\byte\per\second} (which corresponds to the throughput of our SSD performing random writes).

\textbf{Batching effect.}
\label{sec:batching}
Since a call to \texttt{fsync} is costly, the cleanup thread uses batching to avoid calling \texttt{fsync} for each write in the log.
Technically, the cleanup thread consumes entries by batches of a given size. If the log does not contain enough entries, the cleanup thread does not propagate them and waits.
In this experiment, we analyze how \sys reacts to different batch sizes by running a workload intensively writing \SI{20}{\gibi\byte} at random locations.
We use a log of \SI{8}{\gibi\byte}, in order to observe how \sys reacts to batch sizes before and after the saturation.

\Cref{fig:all:fio-batchsize} reports the instantaneous throughput, the average latency and cumulative data written during the run with these settings.
As in \S\ref{eval:log-saturatation}, we can first observe the log saturation at \SI{18}{\second}.
We can also observe that, before saturation, changing the batch size does not affect performance (throughput, latency and written data).
After saturation, we can observe the batch size influence on performance.
With a very small batch size, the throughput is as low as \SI{21}{\mebi\byte\per\second}: for each write, \sys triggers a \texttt{fsync}, which makes \sys less efficient than the SSD configured with \texttt{O\_DIRECT} because of the cost of the system call.
After the saturation, we observe that \sys becomes more efficient with larger batch sizes, for two reasons:
First, \sys does not call \texttt{fsync} often, which increases its performance.
Then, because \sys first writes a batch in the volatile kernel page cache, writes that modify the same location in the file are combined in volatile memory, decreasing the number of pages Linux has to propagate to the SSD upon \texttt{fsync}\added[id=]{\cite{love:13:linuxioscheduler}}.
We also observe that the difference between a batch size of 100, 1000 and 5000 remains low.
This result shows that, as soon as the batch size becomes large enough, the influence of this parameter is low.

\begin{figure*}
    \centering
    \includegraphics{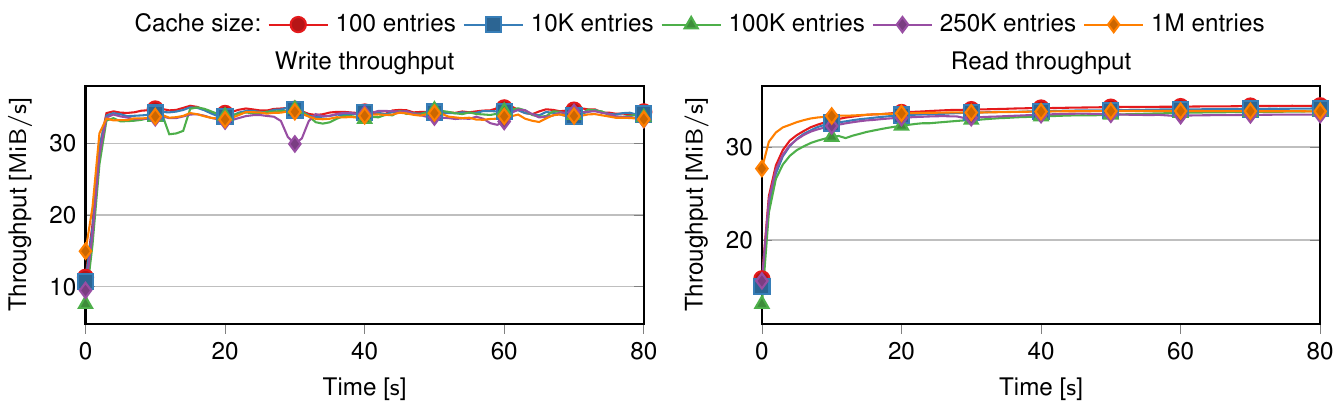}
    \caption{\label{fig:all:fio-readcache} \sys with different read cache size, under mixed read/write loads.}
\end{figure*}

\textbf{Read cache size effect.}
As presented in \S\ref{sec:approach:overview}, since the Linux page cache becomes stale when a pending write in the log is not propagated, \sys relies on a small read cache to ensure consistency.
In this experiment, we show how \sys reacts to different read cache sizes.
\Cref{fig:all:fio-readcache} presents the write (left) and read (right) throughput of \sys with a random read/write workload (50\%/50\%)
\added[id=, comment=]{and a file of \SI{10}{\gibi\byte}.
  Since the read cache is required for consistency, we cannot deactivate it. We start thus
  with a very small cache of 100 entries (\SI{400}{\kibi\byte}) for which the probability of cache hit is negligible.
  The results confirm how the size of the read cache does not affect performance.
  We can observe that with a large cache of \SI{4}{\gibi\byte} (1M entries) where
  the probability of cache hit reaches 40\%, the performance remains the same.
  This result is expected since \sys relies on the kernel page cache to improve performance.
  \sys only uses the read cache to ensure that in case of dirty read (read of a page with a pending write
  in the log), the read is correct and includes the pending writes.
  Our evaluation shows that, because of this redundancy, the size of the \sys read cache does not influence performance,
  which shows that we can keep the read cache small in \sys.}
\deleted[id=RD, comment=question 4]{The figure shows that the size of the read cache does not affect performance.
  This result is expected as the \sys read cache is redundant with the kernel page cache is not meant to improve performance in the general case: it is only useful for consistency to optimize reads in the rare case when a thread writes at a location in a file and then quickly reads the same location. This scenario is not easily observable with FIO, which can only read and write sequentially (no probability of reading data that was just written) or at random locations (only very low probability).}

\section{Related work}
\label{sec:related}

Non-volatile memory solutions have been extensively studied.
The recent introduction in the market of DRAM-based PMEM (\ie, NVDIMM Optane DC persistent memory modules~\cite{optanedc}) initiated a new stream of research exploiting the benefits of these persistent units, as well as novel commercial offerings~\cite{crump:19:optane}.
Studies highlighted their performance trade-offs~\cite{izraelevitz:19:optane,yang:20:pmconcurrency}, some of the compromises for porting legacy applications to the official Intel PMDK~\cite{wang:18:pmdk}, as well as the substantial engineering efforts to port complex systems such as Redis and memcached~\cite{choi:20:observations} or LevelDB~\cite{lersch:17:nvramcache}).

Interestingly, the impact of non-volatile memories has been evaluated in the context of databases~\cite{arulraj2015let,sqlserver-nvdimm}, also including approaches with flash memories~\cite{kang2012flash} or as direct accelerators for main memory~\cite{matthews:08:inteltm}.
As shown in our evaluation, \sys can transparently boost complex DBMS systems, \eg, SQLite or RocksDB, without changes to their source code.
To achieve this transparency, \sys intercepts and redirects in user-space I/O function calls, similar to what SplitFS~\cite{kadekodi:19:splitfs} does using Quill~\cite{eisner2013quill}, an automatic system call wrapper (similar in essence to what we did in our prototype).

A plethora of file systems and I/O boosters have been designed and implemented for non-volatile memory.
\Cref{table:related} (see \S\ref{sec:intro}) presents an analysis of recent research efforts in the field.
To begin, some of the systems implement a file system only tailored for \ac{NVMM}, either by porting an existing file system to \ac{NVMM}~\cite{sehgal:15:nvmfs}, \eg, \extdax~\cite{corbet2014supporting,wilcox2014add}, or by adapting existing file systems to better leverage \ac{NVMM}, \eg, \nova~\cite{xu2016nova} or \splitfs~\cite{kadekodi:19:splitfs}.
Today, because of high prices,\footnote{According to Google Shop in September 2020, \SI{1}{\giga\byte} of \ac{NVMM} remains roughly 100$\times$ more expensive than \SI{1}{\giga\byte} of SATA \ac{SSD}.} \ac{NVMM} comes with a much smaller capacity than mass storage devices such as SSD or HDD, making its use for large workloads currently unrealistic.
Other systems propose to combine \ac{NVMM} and mass storage to offer larger storage space.
They have either the goal of improve reliability~\cite{hu:99:rapid, baker:92:nvmfs} or of boosting I/O performance~\cite{kwon:17:strata, dmwritecache-icpe19, condit:09:betterio, qiu:13:nvmfs, zheng:19:ziggurat, chen:13:fsmac, liu:19:hasfs, niu:18:xpmfs}.
These systems require either modifications in the kernel (hard to maintain) or in the application (hard to apply), or new interfaces between the kernel and the user space (which increase the attack surface of the kernel).
With \sys, we show that we can boost I/O performance without modifying the kernel or the applications, and without requiring coordination between the kernel and the user space through new kernel interfaces.
\dmwritecache is a Linux kernel project that boosts I/O performance, but without requiring new kernel interfaces.
While most of the other systems can efficiently provide new correctness guarantees such as synchronous durability, our evaluation shows that this is not the case with \dmwritecache because the write cache is implemented behind the kernel page cache.
With \sys we show that, by implementing the write cache on top of the kernel page cache (in user land in our case, but we could also implement the write cache in the upper layers of the kernel), we can both boost write performance and efficiently provide advanced correctness guarantees.

Instead of exploiting \ac{NVMM} for file systems, persistent memory can be used directly using \texttt{load} and \texttt{store} instructions.
\Ac{PTM} libraries use transactions to guarantee a consistent state in the event of a non-corrupting failure~\cite{izraelevitz2016failure}.
Typically they intercept every load and store of the application, using one of the following three techniques to guarantee consistency: redo-log, undo-log or shadow data~\cite{volos2011mnemosyne, ramalhete:19:onefile, pmdk, correia2018romulus}.
Most legacy applications that persist data were designed to write to the file system and modifying those applications to use the \acp{PTM} described previously would require substantial re-engineering.

\section{Conclusion}
\label{sec:conclusion}

\sys is a write cache in \ac{NVMM} that improves the write performance of legacy applications.
Unlike with persistent libraries, \sys allows legacy applications to use \ac{NVMM} without any redesign or code modifications.
Our evaluation shows that \ac{NVMM} performs as well as \extfour configured as a direct access file system, and in most cases on par with \nova, but without restricting the storage space to the \ac{NVMM} size.

\medskip\noindent\textbf{Experimental reproducibility.}
We encourage experimental reproducibility.
\added[id=]{The code is freely available at https://github.com/Xarboule/nvcache.}

\section*{Acknowledgments}
This work received funds from the Swiss National Science Foundation (FNS) under project PersiST (no. 178822).

\bibliographystyle{IEEEtranS}
\bibliography{references}

\end{document}